\documentclass{article}

    \PassOptionsToPackage{numbers, compress}{natbib}
\usepackage[preprint]{neurips_2025}

\usepackage[utf8]{inputenc} % allow utf-8 input
\usepackage[T1]{fontenc}    % use 8-bit T1 fonts
\usepackage{hyperref}       % hyperlinks
\usepackage{url}            % simple URL typesetting
\usepackage{booktabs}       % professional-quality tables
\usepackage{amsfonts}       % blackboard math symbols
\usepackage{nicefrac}       % compact symbols for 1/2, etc.
\usepackage{microtype}      % microtypography
\usepackage{xcolor}         % colors
\usepackage{xspace}
\usepackage{graphicx}
\usepackage{wrapfig}
\usepackage{amsmath}
\usepackage{amssymb}
\usepackage{mathtools}
\usepackage{amsthm}
\usepackage{pifont}% http://ctan.org/pkg/pifont
\newcommand{\cmark}{\ding{51}}%
\newcommand{\xmark}{\ding{55}}%

\newcommand{\modelname}{\textsc{TalkPlay}\xspace}
\usepackage{listings}

\lstdefinestyle{json}{
    basicstyle=\ttfamily\small,
    numbers=none,
    breaklines=true,
    morestring=[b]",  % handles strings in double quotes
    stringstyle=\color{green!50!black},
    showstringspaces=false,
    tabsize=2,
    frame=single,
    backgroundcolor=\color{gray!5},
    moredelim=**[is][\color{blue}]{@}{@},  % custom delimiters for specific highlighting
    moredelim=**[is][\color{red}]{|}{|},   % if you need another color for some elements
    captionpos=b
}

\lstdefinestyle{docstring}{
    basicstyle=\ttfamily\small,
    breaklines=true,
    frame=single,
    backgroundcolor=\color{gray!5},
    numbers=none,
    columns=flexible,
    keepspaces=true
}

\title{\modelname: Multimodal Music Recommendation with Large Language Models}
\author{%
  Seungheon Doh~$^{1,2,}$\thanks{Equal contribution. Corresponding authors: seungheondoh@kaist.ac.kr, juhan.nam@kaist.ac.kr;}~,\quad Keunwoo Choi~$^{1,2,*}$,\quad Juhan Nam~$^{1}$ \\
  $^{1}$ KAIST, South Korea, \quad $^{2}$ talkpl.ai, USA \\
  \color{magenta}{\url{https://talkpl-ai.github.io/talkplay}}
}

\begin{document}
\maketitle

\begin{abstract}
We present \modelname, a novel multimodal music recommendation system that reformulates recommendation as a token generation problem using large language models (LLMs). By leveraging the instruction-following and natural language generation capabilities of LLMs, our system effectively recommends music from diverse user queries while generating contextually relevant responses. While pretrained LLMs are primarily designed for text modality, \modelname extends their scope through two key innovations: a \textit{multimodal music tokenizer} that encodes audio features, lyrics, metadata, semantic tags, and playlist co-occurrence signals; and a \textit{vocabulary expansion} mechanism that enables unified processing and generation of both linguistic and music-relevant tokens. By integrating the recommendation system directly into the LLM architecture, \modelname transforms conventional systems by: (1) unifying previous two-stage conversational recommendation systems (recommendation engines and dialogue managers) into a cohesive end-to-end system, (2) effectively utilizing long conversational context for recommendation while maintaining strong performance in extended multi-turn interactions, and (3) generating natural language responses for seamless user interaction. Our qualitative and quantitative evaluation demonstrates that \modelname significantly outperforms unimodal approaches based solely on text or listening history in both recommendation performance and conversational naturalness.
\end{abstract}

\section{Introduction}
\label{intro}

The explosive growth of music streaming platforms has created overwhelming choices for users, with millions of tracks available~\footnote{https://newsroom.spotify.com/company-info/}. Traditional recommendation systems~\cite{hu2008collaborative, koren2009matrix, su2009survey, celma2010music} rely heavily on listening history rather than explicit user queries, while text-to-music retrieval systems~\cite{wu2023music, huang2022mulan, manco2022contrastive, doh2023toward, doh2024enriching} only support single-turn interactions, creating limitations when users are dissatisfied with initial results and need to refine their requests. Conversational recommendation systems (CRS)~\cite{goker2000adaptive, christakopoulou2016towards, zhang2018towards, chen2019towards, jannach2021survey} address these limitations by enabling natural language interactions for item discovery, capturing user preferences, providing relevant recommendations, and generating responses.

In parallel with these developments, recent advancements in large language models (LLMs)~\cite{achiam2023gpt, team2024gemini, grattafiori2024llama} have demonstrated remarkable capabilities across diverse domains, from natural language understanding to complex reasoning tasks. Their ability to follow instructions, generate contextually relevant responses, and leverage extensive pre-trained knowledge has led to increasing exploration of LLMs in recommendation systems~\cite{he2023large, liu2023chatgpt}. However, a key challenge is the semantic gap between LLMs and recommender systems, as items are typically indexed by unique identifiers outside the LLM's vocabulary~\cite{hua2023tutorial}, making direct application difficult. More recent approaches address this by fine-tuning LLMs on recommendation data~\cite{zheng2024adapting, zhang2023recommendation, geng2023vip5}, enabling them to learn associations between natural language queries and item representations. However, the seamless integration of multimodal content with LLMs remains largely unexplored.

To address this problem, we present \modelname, a multimodal music recommendation system powered by LLMs. Our approach reformulates recommendation as next token prediction, also known as generative retrieval~\cite{tay2022transformer, wang2022neural, rajput2023recommender}, through novel \textit{music tokenization} and \textit{vocabulary expansion} techniques. Our system represents music items as token sequences that encode audio features, lyrics, metadata, tags, and playlist co-occurrence, enabling direct recommendations through next-token prediction within conversations. This integrated approach offers several advantages: end-to-end learning without explicit dialogue management, inheritance of broad musical knowledge from pre-trained LLM weights, and effective utilization of multiple information sources (user preferences, queries, interaction patterns) through the LLM architecture.

Our experimental results demonstrate that: (1) the proposed multimodal recommendation system outperforms unimodal approaches based solely on text or listening history; (2) our LLM-based generative retrieval system maintains strong recommendation performance even in extended multi-turn contexts; and (3) compared to two-stage systems, our approach shows superior performance in human evaluations of recommendation relevance and response generation quality.

\begin{figure*}[!t]
\centering
\includegraphics[width= \linewidth]{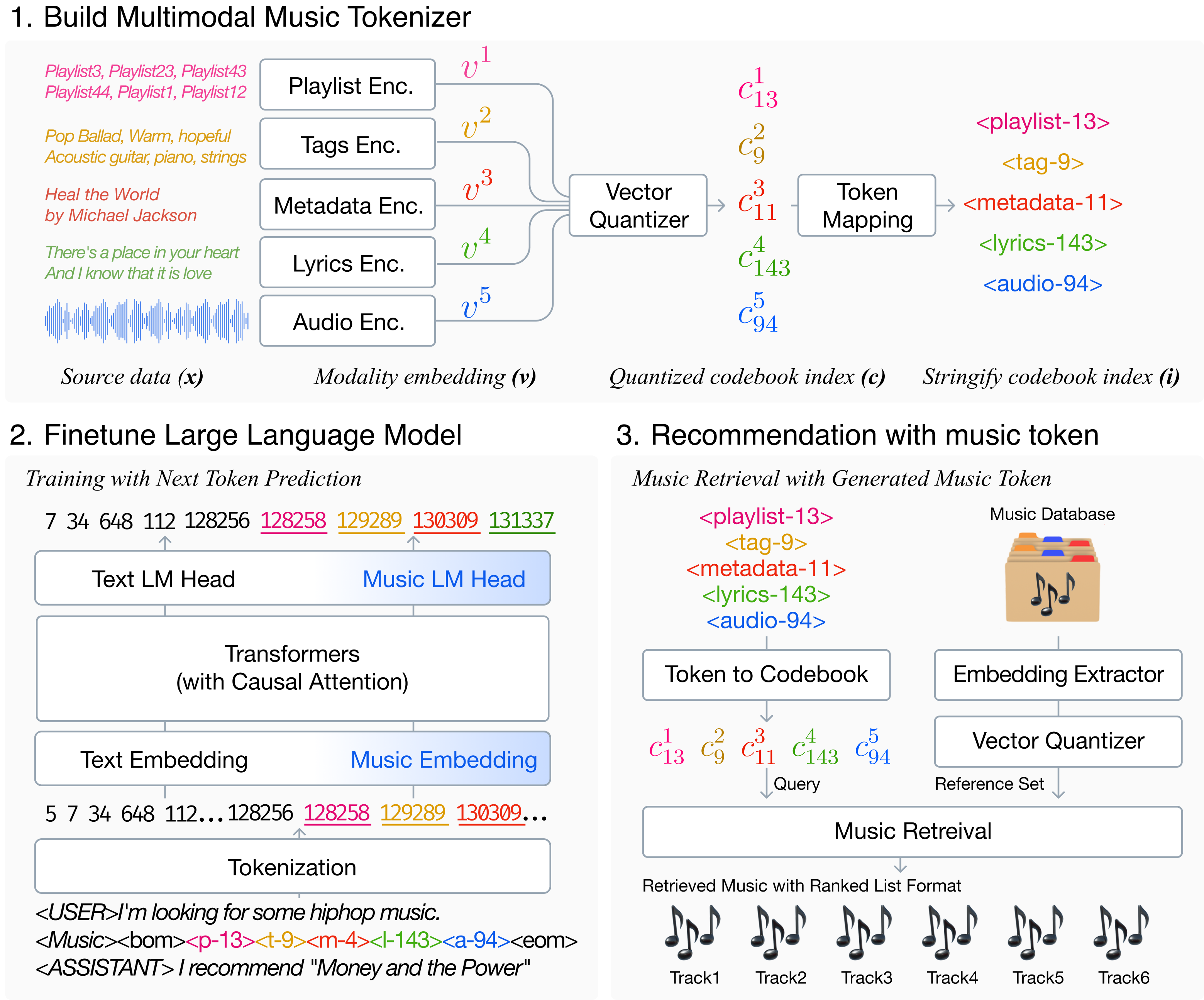}
\caption{Overview of \modelname: (1) The multimodal music tokenizer converts source data into modality embeddings $v$ and quantizes them into codebook indices $c$, which are mapped to music tokens $i$. (2) The LLM is fine-tuned on text and music token sequences. (3) The generated music tokens are mapped back to codebook indices and are used as queries to retrieve music items from the database.}
\label{fig:talkplay-overview}
\end{figure*}

\section{\modelname: The Proposed System}\label{sec:proposed-system}

As illustrated in Figure~\ref{fig:talkplay-overview}, \modelname consists of two parts: a music tokenizer and a large language model. The music tokenizer encodes multimodal music embeddings into tokens. A single music item is represented by multiple tokens, each of which represents a different modality. The model is a multimodal LLM that is presumably: i) capable of modeling music recommendation conversations (with text and music tokens), and ii) pretrained with a large amount of data including musical corpus.

\subsection{Music Tokenizer}\label{subsec:system-music-token}
The music tokenizer in \modelname is designed under following goals and constraints. First, to retrieve and recommend items, the model needs to be connected to the database. Second, to easily utilize LLM as a recommendation engine, the items in the database need to be discretized as tokens. Third, the embeddings of those item tokens need to be trained well, requiring some minimum token occurrence in the training data. Given these conditions, we propose the following scheme for music tokenization in \modelname. A music item $x_i$ ($i$ is an item index) may have rich information including (1) playlist co-occurrence, (2) semantic tag information, (3) metadata, (4) lyrics, and (5) audio. 

\textbf{Item Embeddings}: For all the data modalities, pretrained models are used to extract item embedding $v_i^m$, where $m$ stands for a modality index (1-5 for playlist co-occurrence, semantic tag, metadata, lyrics, and audio, respectively.) For playlist co-occurrence, a word2vec-style~\cite{barkan2016item2vec} item embedding model is trained using the training dataset ($v_i^1 \in \mathbb{R}^{128}$). For semantic information, metadata, and lyrics, NV-Embed-v2~\cite{lee2024nv} is chosen as its state-of-the-art performance in various text embedding benchmarks~\cite{muennighoff2022mteb} ($v_i^2, v_i^3, v_i^4 \in \mathbb{R}^{4096}$).  For audio, the MusicFM~\cite{won2024foundation} model is chosen for its strong performance in many music informatics tasks ($v_i^5 \in \mathbb{R}^{1024}$). 

\textbf{Quantization}: Due to the long-tail nature of music consumption and the limited size of available training data, all the item embeddings $v_i^m$ are vector-quantized into $K$ clusters (per modality) using K-means clustering. Here, $c_k^m$ represents a cluster centroid. In this paper, we chose $K=1024$.
As a result, the overall music database is represented by $K \times M$ indices, where the total number of modalities $M$ is 5. However, this does not mean that all music items are reduced to $1024 \times 5=5120$ unique clusters. A music item is represented by 5 modality tokens, where each modality embedding has 1024 clusters. Theoretically, the proposed music tokenization scheme is capable of representing $1024 ^ 5$ = about 1,126 trillion unique music items.

\textbf{Token Mapping}: In this final step of tokenization step, the multimodal cluster indices of each music item is used in our token format: \texttt{<|modality-index|>}. For example, a music item whose embeddings are mapped to 29th audio cluster, 2nd metadata cluster, 98th lyrics cluster, 3th semantic cluster, and 59th user-item cluster is converted to \texttt{<|playlist-59|>}\texttt{<|semantic-361|>}
\texttt{<|metadata-7|>}\texttt{<|lyrics-98|>}\texttt{<|audio-29|>}. 

\subsection{LLM: The Recommender}\label{subsec:system-llm}
The LLM, our recommendation engine, is based on a pretrained model. We opt for instruction finetuned model to utilize the conversationability as our goal is to build a conversational recommender. Our task is highly specific compared to the scope of usual LLM training, implying relatively small models might be enough. Altogether, Llama-3.2-1B-Instruct is chosen for our base model~\cite{dubey2024llama}. 

To enhance the LLM's understanding of multimodal music information, we expand its vocabulary to include multimodal music tokens~\cite{zhao2024llama}. In detail, the vocabulary size of the base model is expanded by $1024 \times 5=5120$, to include the multimodal music tokens. Two more special tokens, \texttt{<start\_of\_music>} and \texttt{<end\_of\_music>}, are added as well, since they are used to wrap music tokens in the training data. This modification means 5122 new embeddings ($e_k^m$ and two special token embeddings) to the model, each of which is 2048-dim in the Llama-3.2-1B model. Equivalently, it adds about 10.5M parameters to the model, increasing the model size by less than 1\%. 
In the synthetic conversation, music items are represented with the multimodal music tokens, which are tokenized using the added vocabulary and fed to the model. This allows the model to directly understand and generate multimodal music tokens along with regular text tokens during conversations.

% To enhance the LLM's understanding of multimodal music information, we expand its vocabulary to include multimodal music tokens~\cite{zhao2024llama}. 
% % 
% In detail, the vocabulary size of the base model is expanded by $1024 \times 5=5120$, to include the multimodal music tokens. Two more special tokens, \texttt{<start\_of\_music>} and \texttt{<end\_of\_music>}, are added as well, since they are used to wrap music tokens in the training data. This modification means 5122 new embeddings ($e_k^m$ and two special token embeddings) to the model, each of which is 2048-dim in the Llama-3.2-1B model. Equivalently, it adds about 10.5M parameters to the model, increasing the model size by less than 1\%. With this vocabulary expansion, in the synthetic conversation, music items are represented with the multimodal music tokens, which are then tokenized and fed to the model. This expansion allows the model to directly understand and generate multimodal music tokens along with regular text tokens during conversations.

\subsection{Training}\label{subsec:system-training}

\modelname employs a principled supervised finetuning approach for LLMs, leveraging autoregressive next token prediction with cross-entropy loss. We transform synthetic conversational data (Section~\ref{subsec:maintask-synthe-music-convo}) by systematically replacing track identifiers with their corresponding multimodal music tokens. Formally, given a token sequence $x = \{(q_1, m_1, r_1), ..., (q_t, m_t, r_t)\}$ representing a conversation with user queries $q$, recommended music tokens $m$, and system responses $r$, the model is trained to predict each subsequent token in the sequence. Each token $t$ can be either a text token from user queries or system responses, or a music token from recommended items. We use the standard autoregressive cross-entropy loss: $\mathcal{L} = -\frac{1}{n} \sum_{i=1}^{n} \log P(t_i | t_1, t_2, ..., t_{i-1})$, where $P(t_i | t_1, t_2, ..., t_{i-1})$ represents the model's predicted probability of token $t_i$ given its preceding context.

This unified formulation enables the model to first predict contextually appropriate music tokens based on the user query $p(m|q)$, and subsequently generate coherent responses conditioned on both the query and predicted music tokens $p(r|q,m)$. Building on insights from prior research~\cite{deldjoo2024content}, we leverage the inherent coarse-to-fine structure of music modalities by predicting tokens in a carefully chosen sequence: \textit{playlist - semantic - metadata - lyrics - audio} tokens.  We validate this ordering strategy through comprehensive ablation studies on feature importance (Section~\ref{results:feature}).

The central goal of this approach is to effectively learn the newly introduced multimodal music token embeddings $e_k^m$ by capturing rich contextual relationships between conversations and music items. We initialize these music token embeddings using the statistics of the base model's existing embeddings, following the vocabulary expansion method proposed by Hewitt et al.~\cite{hewitt2021initializing}, $e_k^m \sim \mathcal{N}(\mu_E, \Sigma_E)$, where $\mu_E$ and $\Sigma_E$ represent the mean and covariance of the base model's token embeddings. 
% This training methodology enables the model to generate contextually relevant multimodal music tokens for recommendations while maintaining fluent and coherent natural language responses.

\subsection{Inference: Recommendation by Generation} \label{subsec:system-inference}
During inference, we use typical LLM generation parameters (temperature=1.0, top-p=0.9, repetition penalty=1.0). The music tokens are generated before the assistant's response message to ensure recommendations are based solely on the query, avoiding potential popularity biases that could arise from conditioning on generated text \cite{he2023large}.

The generated tokens must then be mapped back to music items through a reverse lookup of the tokenization process described in Section~\ref{subsec:system-music-token}. While each individual cluster codebook index has a many-to-one track-to-cluster relationship, our tokenizer creates sequences of five multimodal tokens that form a nearly one-to-one track-to-cluster-sequence mapping (similar to Semantic IDs~\cite{rajput2023recommender} using residual vector quantization), making the process more deterministic than typical approximate nearest neighbor searches used in industry recommenders \cite{spotify_annoy}.

Given a sequence of five multimodal tokens representing a recommendation, we select items matching the token indices across modalities. However, due to the vast token space (1,126 trillion combinations) relative to our dataset size, exact matches are often unavailable. In such cases, we fall back to partial matching using token overlap scoring with feature importance weights assigned to each modality. The final score is determined as $s_\text{retrieval} = \sum_{i \in \{\text{modalities}\}} \lambda_i s_i$, where each $s_i$ term is a binary indicator (0 or 1) representing whether there is an overlap in the corresponding modality. Based on our preliminary experiment, we use a quadratically decreasing weighting scheme that best captures the relative importance of different music features. The final weights used are $\lambda_\text{cf}=25$, $\lambda_\text{semantic}=16$, $\lambda_\text{metadata}=9$, $\lambda_\text{lyrics}=4$, and $\lambda_\text{audio}=1$. This weighting scheme reflects our finding that playlist co-occurrence token provide the strongest recommendation performance gain, followed by semantic tags, metadata, lyrics, and audio features. The results of these weighting experiments are presented in detail in Section~\ref{results:feature}.

\section{Data Generation}\label{sec:data}
It is prohibitively difficult to perform research on recommendation without access to real-world data -- information about items, users, and their interactions. Another challenge lies in the fact that the conversational recommendation system is in its early stage and so the publicly available dataset is scarce. The Conversational Playlist Curation Dataset (CPCD) is probably the most relevant one, but it comes with 917 conversations which are not sufficient to train a large model~\cite{chaganty2023beyond}.

Inspired by recent works such as ~\cite{leszczynski2023talk, doh2024music}, we generate the training data using a music playlist dataset and an LLM. These synthetic datasets have been shown to closely resemble human conversations in qualitative evaluations and have proven effective for training recommendation models. Following our primary reference ~\cite{doh2024music}, each conversation is generated as a journey of machine-assisted music discovery; assistants continue to react by choosing music items that satisfy the user's request. A notable difference is that in \modelname, the base dataset is not simply a music dataset (e.g., \cite{bertin2011million}) but a playlist dataset, in order to simulate more plausible and coherent conversations and recommendations.

\begin{wraptable}{r}{0.5\textwidth}
\vspace{-13.8mm}
\caption{Statistics of the \modelname dataset.}
\vspace{2mm}
\label{tab:dataset-stats}
\centering
\resizebox{0.49\textwidth}{!}{
\begin{tabular}{lrr}
\toprule
Dataset                       & Training         & Evaluation       \\ \midrule
\multicolumn{3}{l}{{\color[HTML]{9B9B9B} \textit{Million Playlist Dataset (MPD)}}} \\
\# of Playlists                & 999,000          & 1,000            \\
\# of Average Tracklist        & 63.73            & 59.41            \\
\# of Warm Tracks        & 1,714,772        & 37,114           \\
\# of Cold Tracks        & -                & 2,357            \\
\# of Vocab                    & 24,129            & 713              \\
\midrule
\multicolumn{3}{l}{{\color[HTML]{9B9B9B} \textit{TalkPlay Dataset}}}        \\
\# of Playlist                 & 116,099          & 1,000             \\
\# of Dialogue                 & 531,552          & 1,000             \\
\# of Tracks                   & 405,543           & 11,006            \\
\# of Vocab                    & 317,705           & 19,736            \\
Avg. \# of turns               & 6.95             & 13.05            \\
% Avg. query len.               & 17.62            & 15.35           \\
% Avg. reponse len.               & 17.91            & 14.92           \\
\bottomrule
\end{tabular}
}
\vspace{-5mm}
\end{wraptable}

\subsection{Base Dataset: Million Playlist Dataset}\label{subsec:base-data}
In the current era of music streaming, playlists are playing a crucial role in the music listening experience, providing unique and direct information about music items: which music items go along together well, according to the real listeners. They may suit well for the purpose of training \modelname as well, as a base dataset, since conversation consisting of tracks in the same playlist would be somewhat diverse yet coherent, and hence realistic. % than other hypothetical strategies such as random sampling or any other heuristics. 

The Million Playlist Dataset (MPD) is chosen as our base dataset \cite{chen2018recsys}. Since being introduced in 2018, this dataset remains one of the largest playlist datasets available to the public by including one million Spotify playlists. The MPD provides track metadata and playlist co-occurrence and is extended with audio content, lyrics, and semantic annotations. The audio content are successfully crawled through the Spotify API for 1,717,148 items. The lyrics are extracted using \textsc{Whisper-Large-V3}~\cite{radford2023robust}, a well-known speech transcription model that performs decent lyrics transcription. For semantic annotations, we leverage pretrained music understanding models to extract comprehensive musical attributes. Specifically, we use \textsc{LP-MusicCaps}~\cite{doh2023lp} to extract high-level semantic features including genres, moods, instruments, and vocal styles. Additionally, we employ \textsc{Madmom}~\cite{bock2016madmom} to extract structural music properties such as key signatures and tempo information. This multi-faceted approach ensures our model has access to a rich representation of musical characteristics.

\subsection{TalkPlayData-1}\label{subsec:maintask-synthe-music-convo}

We use an LLM and the base dataset (MPD) to generate TalkPlayData-1, a music conversation dataset with natural language and music recommendation.\footnote{https://huggingface.co/collections/talkpl-ai/talkplay-data-1} These are the key design principles and choices to ensure the synthetic data can be used for the training of \modelname.

\textbf{Coherence}: Given a user query/reaction, system should recommend music with relevant dialogue.

\textbf{Modality Coverage}: Conversations span four target modalities: audio, lyrics, metadata, and semantic tags. The playlist co-occurrence is excluded because: i) this modality is implicit and difficult to verbalize, and ii) in our task setup, playlist co-occurrence would be inherently captured (Section~\ref{subsec:system-music-token}).

\textbf{User Simulation}: The user may skip or reject the recommended tracks, requesting for alternative recommendations. For rejected tracks, we employ hard negative sampling by using songs from artists whose tracks appear in the playlist. This approach ensures that the rejected recommendations are plausible but not suitable for the current conversation context.

\textbf{Text Format}: The LLM output, the synthesized conversation, should be in a structured format. The JSON format is chosen and used as presented in Section~\ref{subapp:synth-convo}

Based on this, \textsc{Gemini-1.5-flash-002}~\cite{team2024gemini} is used to create music conversation given a prompt that includes i) our goal and the aforementioned conditions, ii) track identifiers and their metadata, semantic tags, lyrics. The prompt is included in Section~ \ref{subapp:synth-prompt}. Table~\ref{tab:dataset-stats} compares our dataset with the base dataset. Using 116k playlists with an average of 7 tracks each, our synthesis approach yields 532k multi-turn music recommendation dialogues. By leveraging rich textual annotations of music sequences, our LLM-based data synthesis technique significantly expands the text-music associations beyond just playlist titles, increasing the token count from 24k to 318k. Each user-assistant interaction in our synthetic conversation dataset follows a triplet structure: \{\textsc{user query, recommended music, assistant response}\}. The user query and assistant response are natural language expressions, while the recommended music is represented by a unique track identifier.

% During training, both the assistant response and the recommended music are used as labels. 
% This structured format resembles the typical task or chat templates. 

\textbf{Data Split}: To reflect the real-world recommendation scenarios, we perform a chronological data split based on the playlist creation date. The test set is created by randomly sampling 1,000 playlists from the last creation date (2017-11-01), and training set include all the rest of the tracks. Through this split, around 2,357 tracks in the test split are cold-start items, tracks that only appear in the test set. Assuming a real-world scenario, these items represent newly released songs, which traditional collaborative filtering algorithms have no information to make recommendation from. In contrast, being multimodal, \modelname is able to handle these items by understanding content such as audio, lyrics, metadata and semantic features.

\begin{wraptable}{r}{0.5\textwidth}
\vspace{-6mm}
\caption{Human evaluation results comparing the quality of conversational datasets.}
\vspace{1mm}
\label{tab:human_eval}
\centering
\begin{tabular}{lcc}
\toprule
Dataset & Relevance & Naturalness \\ \midrule
CPCD & 4.08 ± 0.11 & 4.01 ± 0.11 \\
LP-MusicDialog & 3.90 ± 0.10 & 3.95 ± 0.11 \\
TalkPlayData-1 & 4.04 ± 0.10 & 4.01 ± 0.09 \\
\bottomrule
\vspace{-7mm}
\end{tabular}
\end{wraptable}

\textbf{Human Evaluation: }~\label{sec:human_eval} We assess the quality of our synthetic conversational dataset through a human evaluation, focusing on two key aspects: (1) \textit{relevance} -- the alignment between the retrieved music items and the user query, and (2) \textit{naturalness} -- the likelihood that such a conversation would occur in real life. We adopt a mean opinion score with a 5-point Likert scale. A total of 26 raters each evaluated 10 randomly sampled dialogues, resulting in 260 ratings in total. For comparison, we include open-source conversational music recommendation datasets: CPCD~\cite{chaganty2023beyond}, a human conversation dataset, and LP-MusicDialog~\cite{doh2024music}, a synthetic conversation dataset. As shown in Table~\ref{tab:human_eval}, TalkPlayData-1 achieves comparable quality to the human-annotated CPCD dataset, with scores of 4.04 and 4.01 for relevance and naturalness respectively. This demonstrates that our LLM-based synthesis approach can generate conversations that are on par with human-created dialogues while being more scalable.

\vspace{-2mm}
\section{Experiments}
\vspace{-2mm}
We conduct a comprehensive evaluation of \modelname, focusing on two critical dimensions: (1) retrieval performance - the model's ability to recommend relevant music items based on multi-turn dialogue context, and (2) response quality - the model's capability to generate contextually appropriate, natural language responses that explain recommendations. Through both objective metrics and human evaluation, we examine how our generative approach compares to traditional retrieval methods across various conversation scenarios and user preferences.

\textbf{Objective Evaluation:} For the multi-turn recommendation evaluation, we simulate a turn-taking scenario where each turn is conditioned on the previous turns and the current turn's user query, with the goal of retrieving appropriate music recommendations. Following previous works~\cite{chaganty2023beyond, doh2024music}, we employ standard ranking metrics - Mean Reciprocal Rank (MRR) and Hit Rate at K (Hit@K). We evaluate retrieval performance for each turn using 1,000 multi-turn conversations derived from 1,000 playlists in Table~\ref{tab:dataset-stats}.
% MRR measures the average reciprocal of the rank at which the first relevant item appears, while Hit@K indicates whether a relevant item 
% appears in the top-K retrieved results. 

\textbf{Subjective Evaluation:} TTo assess the real-world usability of \modelname, we conducted rigorous A-vs-B human preference evaluations against baseline models and ground truth. In each evaluation scenario, evaluators compared a pair of outputs (recommendations and responses) from different models based on a shared conversation context, to choose a preferred model (or a tie) on the two critical dimensions: (1) \textit{Recommendation Relevance}, and (2) \textit{Response Naturalness} as mentioned in Section~\ref{sec:human_eval}. We recruited 18 evaluators with expertise in music research or professional experience in the music industry. Each evaluator assessed 10 randomly selected conversation scenarios, yielding 180 total ratings.

\textbf{Training Details:} Using the synthetic conversation dataset (Section~\ref{sec:data}), \modelname, a vocabulary-expanded Llama-3.2-1B model, is trained with a 1024-token context length, a learning rate of $1e-4$, and the AdamW optimizer \cite{loshchilov2017decoupled}. The model reported in this paper is trained on 8 GPUs with a total batch size of 48 items (or 49,152 tokens).

\textbf{Baseline Models:} We evaluate our model against three baseline approaches: BM25~\cite{robertson1999okapi} for classical text-based sparse retrieval, NV-Embeds-V2~\cite{lee2024nv} for deep similarity retrieval with LLM embeddings, and CLAP~\cite{wu2023large} for multimodal audio-text retrieval. We also include two generative retrieval models, SASRec~\cite{kang2018self} and TIGER~\cite{rajput2023recommender}. While both models perform next token prediction for recommendation, SASRec represents each of the 400k tracks as unique IDs, while TIGER uses Semantic IDs that combine SentenceT5~\cite{ni2021sentence} embeddings with residual vector quantization (RVQ) for track representation. Unlike \modelname, SASRec and TIGER lack natural language understanding capabilities. As a result, these models can only make recommendations based on track sequences, ignoring the semantic content of user queries in multi-turn conversations. Additional details can be found in Appendix~\ref{appendix_a_4}.

For our subjective evaluation, we chose BM25~\cite{robertson1999okapi} and NV-Embeds~\cite{lee2024nv} as baseline models, as they demonstrated strong query understanding capabilities in our objective evaluation (In Table~\ref{tab:retrieval-results}). To enable these retrieval-based models to generate conversational responses, we implemented a two-stage CRS using Llama-3.2 1B~\cite{dubey2024llama}. This pipeline first retrieves relevant music items using retrieval model, then uses these retrieved items as context for the LLM to generate appropriate responses.

\begin{table}[!t]
\caption{Comparison of recommendation performance across different models. MRR indicates the mean reciprocal rank and Hit@K measures proportion of relevant items in top-K results. QU and RG indicate query understanding and response generation capabilities, respectively.}
\label{tab:retrieval-results}
\centering
\begin{tabular}{llccrrr}
\toprule
Models & Retrieval Methods & QU & RG & MRR($\uparrow$) & Hit@1($\uparrow$) & Hit@100($\uparrow$) \\ \midrule
\multicolumn{7}{l}{{\color[HTML]{9B9B9B} \textit{\small{Zero-shot Recommendation Models}}}} \\
BM25~\cite{robertson1999okapi}  & Sparse & \textcolor{green!70!black}{\cmark} & \textcolor{red}{\xmark} & 0.030 & 0.002 & 0.265 \\
CLAP-Music~\cite{wu2023large}  & Similarity-based & \textcolor{green!70!black}{\cmark} & \textcolor{red}{\xmark} & 0.011 & 0.003 & 0.153 \\
NV-Embed-V2~\cite{lee2024nv} & Similarity-based & \textcolor{green!70!black}{\cmark} & \textcolor{red}{\xmark} & 0.029 & 0.005 & 0.255 \\ \midrule
\multicolumn{7}{l}{{\color[HTML]{9B9B9B} \textit{\small{Finetuned Recommendation Models}}}} \\
CLAP-Music~\cite{wu2023large} & Similarity-based & \textcolor{green!70!black}{\cmark} & \textcolor{red}{\xmark} & 0.017 & 0.005 & 0.189 \\
NV-Embed-V2~\cite{lee2024nv} & Similarity-based & \textcolor{green!70!black}{\cmark} & \textcolor{red}{\xmark} & 0.032 & 0.005 & 0.281 \\
SASRec~\cite{kang2018self} & Generative & \textcolor{red}{\xmark} & \textcolor{red}{\xmark} & 0.002 & 0.000 & 0.008 \\
TIGER~\cite{rajput2023recommender} & Generative & \textcolor{red}{\xmark} & \textcolor{red}{\xmark} & 0.008 & 0.001 & 0.092 \\
\modelname (Ours) & Generative & \textcolor{green!70!black}{\cmark} & \textcolor{green!70!black}{\cmark} & \textbf{0.049} & \textbf{0.026} & \textbf{0.288} \\ \bottomrule
\end{tabular}
\end{table}

\section{Results}
\label{sec:results}
\subsection{Conversational Music Recommendation Results}
Our results in Table~\ref{tab:retrieval-results} demonstrate that \modelname significantly outperforms both zero-shot and finetuned baseline models across all key metrics. \modelname achieves an MRR of 0.049, which is higher than the best baseline (finetuned NV-Embed-V2 at 0.032). More notably, \modelname shows a dramatic improvement in Hit@1 performance (0.026), achieving more than 5 times the accuracy of the next best model. This demonstrates a strong correlation between the cross-entropy loss function used in next token prediction and the Hit@1 metric. The superior Hit@1 performance of \modelname suggests that our generative approach can directly produce highly relevant recommendations without requiring extensive candidate consideration in the realistic music listening experience.

Among zero-shot recommendation models (upper side of Table~\ref{tab:retrieval-results}), traditional sparse retrieval methods such as BM25 and embedding-based approaches like NV-Embed-V2 exhibit comparable performance, with MRR scores of 0.030 and 0.029 respectively. This indicates that, in the absence of task-specific training, dense embedding method~\cite{lee2024nv} offer limited benefit over simpler retrieval techniques~\cite{robertson1999okapi}. One reason for this close performance is that a large portion of music-related queries involve metadata—such as track titles, artist names, and album information—where sparse retrieval methods like BM25 are particularly effective due to their strength in exact token matching. In contrast, CLAP~\cite{wu2023large} performs significantly worse in multi-turn recommendation scenarios (MRR of 0.011). Given that CLAP was trained primarily on semantic annotations like genre and instruments without conversational context, this suggests its limited dialogue understanding capability leads to lower performance in conversation-based recommendation scenarios.

In fine-tune recommendation models (lower side of Table~\ref{tab:retrieval-results}), fine-tuning improves performance for all baseline models, but the gains are modest compared to \modelname's results. Notably, the sequencial recommendation models, such as SASRec~\cite{kang2018self} and TIGER~\cite{rajput2023recommender}, perform significantly worse than retrieval-based approaches. It indicates that the challenge of generating accurate recommendations without query understanding capabilities in conversational recommendation scenario. 

\begin{wrapfigure}{r}{0.52\textwidth}
\vspace{-10pt}
\centering
\includegraphics[width=0.52\textwidth]{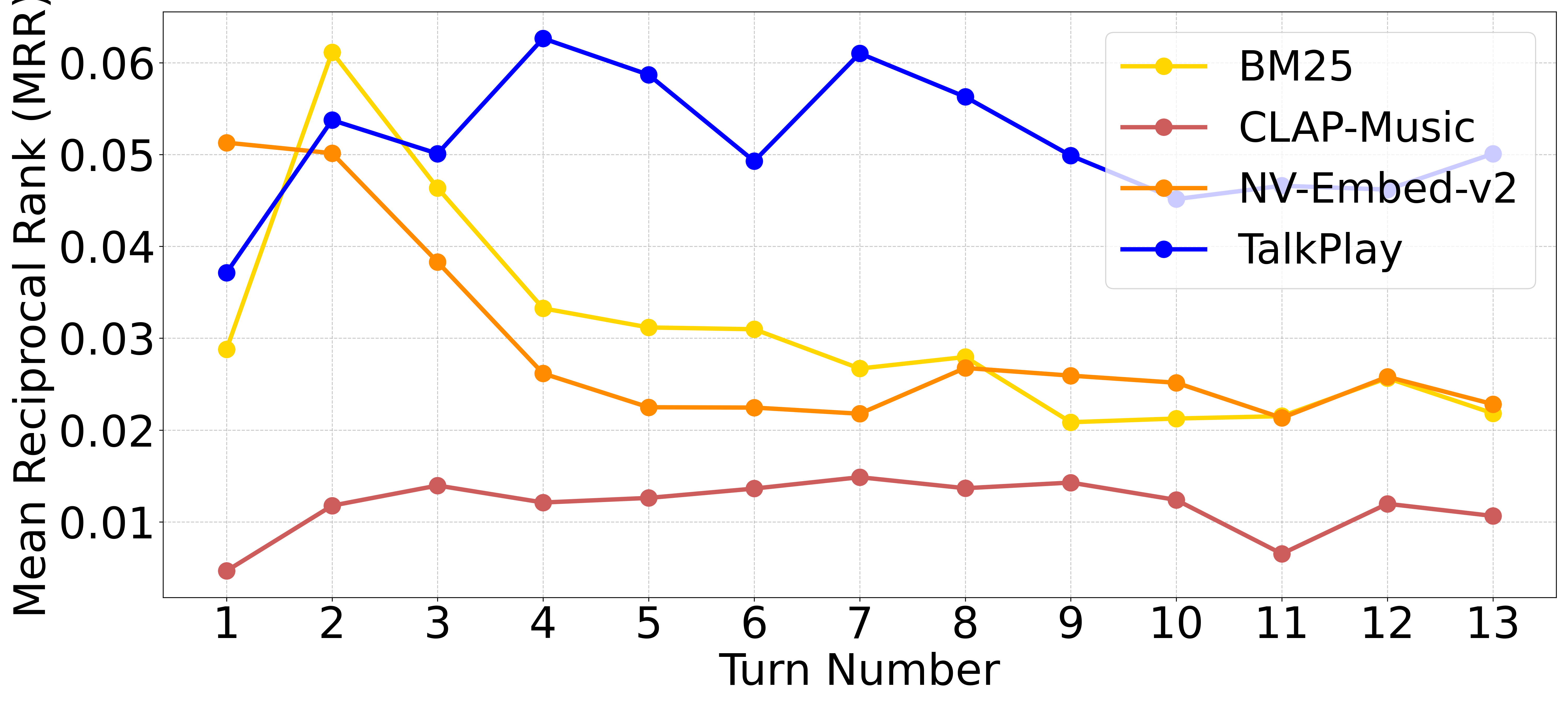}
\vspace{-5mm}
\caption{Performance comparison across conversation turns. The x-axis shows the turn number, and the y-axis shows Mean Reciprocal Rank.}
\label{fig:turn_wise_eval}
\vspace{-3mm}
\end{wrapfigure}

Figure~\ref{fig:turn_wise_eval} shows how retrieval performance varies across conversation turns. The MRR scores reveal an interesting pattern: while \modelname starts with comparatively lower scores in the first 1-2 turns, it significantly outperforms all baselines from Turn 4 onward and maintains relatively stable performance even as the conversation extends to later turns to Turn 13. In contrast, baseline methods exhibit notable degradation as the conversation continues. This pattern shows a fundamental trade-off in conversational recommendation systems: \modelname requires a few dialogue exchanges to establish user preferences and context, but once established, it excels at maintaining this understanding throughout extended interactions. 
% The initial lower performance reflects the inherent challenge of making recommendations with minimal contextual information - a common cold-start scenario in conversational systems. 

% Another notable aspect of multi-turn recommendation is that similarity-based retrieval methods rely on simple aggregation of previous chat history for retrieval, which becomes less effective as context grows more complex. 
This result also emphasizes the fundamental challenge of multi-turn recommendation. Similarity-based retrieval methods must rely on simple aggregation strategies (e.g., averaging) to combine the previous chat history, which becomes less effective as context grows more complex. In contrast, \modelname's LLM-based approach leverages the model's internal attention mechanism during token prediction to naturally capture and prioritize relevant contextual signals throughout the conversation. This enables more nuanced understanding of evolving user preferences, particularly beneficial in real-world music discovery scenarios where conversations typically extend beyond a few initial exchanges.

\subsection{Qualitative Results}

\begin{figure}[!t]
   \centering
   \includegraphics[width=\textwidth]{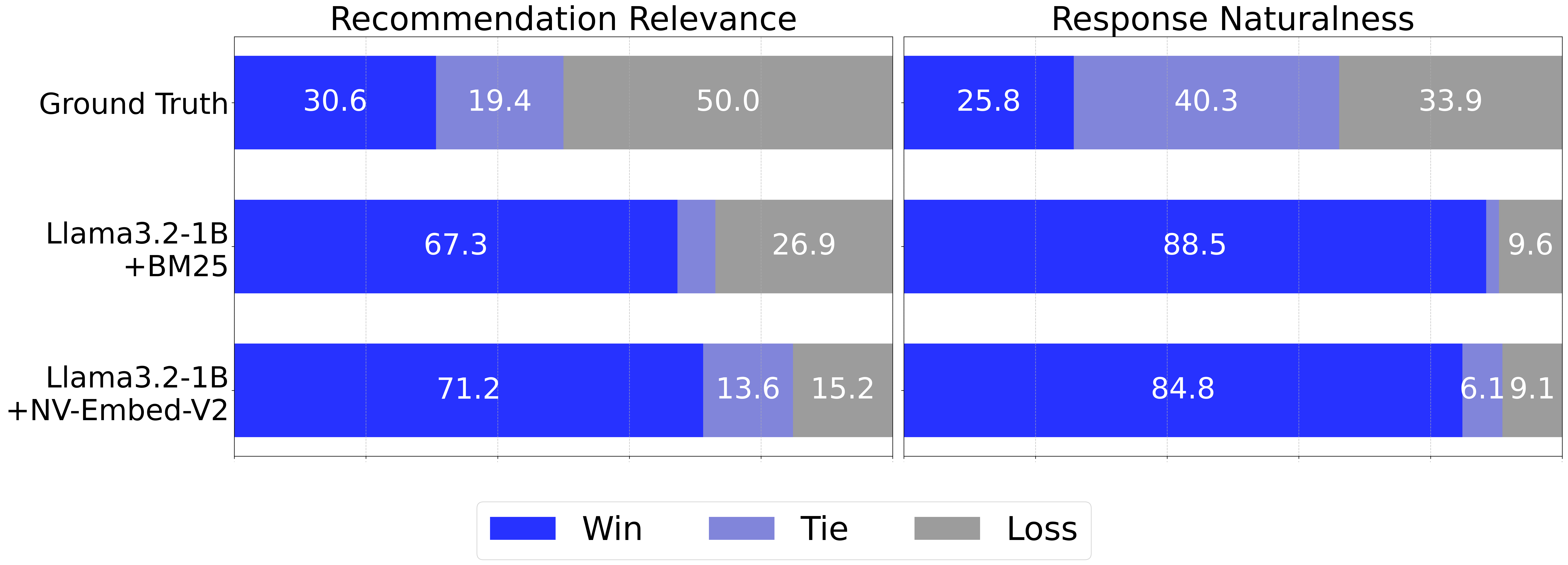}
   \caption{A-vs-B human evaluation results, comparing \modelname against existing conversational music recommendation models on recommendation relevance (left) and response naturalness (right).}
   \label{fig:model_comparison_results}
   \vspace{-5mm}
\end{figure}

Figure~\ref{fig:model_comparison_results} presents the results of our A-vs-B human evaluation, which compares \modelname against both ground truth data and a set of baseline models. We conducted a human evaluation study to assess the quality of our model's recommendations and responses. Participants were presented with multi-turn conversation scenarios and asked to compare recommendations and generated responses between two models. The survey interface is showed in the appendix~\ref{appendix_a_3}.

In the comparison with ground truth data,\modelname demonstrates competitive performance: it achieves a combined \textit{Win+Tie} rate of 50\% for the \textit{Recommendation Relevance} metric and 66.1\% for \textit{Response Naturalness}. These scores indicate that the outputs generated by \modelname are on par with ground truth recommendations and responses in terms of both relevance and fluency.

When compared against baseline models, \modelname exhibits a clear performance advantage. Specifically, in the domain of recommendation relevance, our model outperforms both baselines: Llama3.2-1B + NV-Embeds-V2 with a Win Rate of 71.2\% and Llama3.2-1B + BM25 with a Win Rate of 67.3\%. These results validate the superiority of our LLM-based approach over traditional retrieval-based baselines in multi-turn recommendation scenario.

The improvements are even more substantial in the response generation task, underscoring the proposed model's capability to generate more natural and contextually appropriate replies within music-related conversations. Human evaluators consistently preferred our model's responses, with Win Rates of 84.8\% and 88.5\% over those generated by the general-purpose Llama-3.2-1B Instruct model. 

\subsection{Multimodal Feature Importance} \label{results:feature}

Tables~\ref{tab:ablation-weight} and~\ref{tab:ablation-feature} present comprehensive ablation studies that systematically analyze the relative importance of different modalities and their weighting strategies in our model. Our analysis of weighting strategies in Table~\ref{tab:ablation-weight} reveals a clear pattern: approaches that emphasize coarse-grained features (playlist and semantic tokens) over fine-grained features (lyrics and audio tokens) consistently demonstrate superior performance. Specifically, both linear and quadratic coarse-to-fine weighting schemes achieve an MRR of $0.049$, representing a substantial 29\% improvement over the uniform weighting baseline ($0.038$). This performance gap strongly suggests that higher-level contextual information encoded in playlists and semantic descriptors serves as a more reliable signal for conversational recommendation compared to granular feature-level information.

\begin{table}[!t]
\caption{Ablation study results showing the impact of different weighting strategies for modality tokens in \modelname. The modality weights are applied in the order: (playlist, metadata, attribute, lyrics, audio).}
\label{tab:ablation-weight}
\centering
\begin{tabular}{llrrrr}
\toprule
Weight Methods & Modality Weight & MRR $\uparrow$   & Hit@1 $\uparrow$ & Hit@10 $\uparrow$ & Hit@100 $\uparrow$ \\
\midrule
Uniform & (1, 1, 1, 1, 1) & 0.038 & 0.023 & 0.066 & 0.180 \\
Linear (fine-to-coarse) & (1, 2, 3, 4, 5) & 0.025 & 0.018 & 0.037 & 0.085 \\
Quadratic (fine-to-coarse) & (1, 4, 9, 16, 25) & 0.029 & 0.020 & 0.043 & 0.124 \\
Linear (coarse-to-fine) & (5, 4, 3, 2, 1) & 0.049 & 0.027 & 0.090 & 0.283 \\
Quadratic (coarse-to-fine) & (25, 16, 9, 4, 1) & 0.049 & 0.026 & 0.096 & 0.288 \\
\bottomrule
\end{tabular}
\end{table}

\begin{wraptable}{r}{0.5\textwidth}
\centering
\vspace{-7mm}
\caption{Ablation study results showing the impact of removing different modalities from \modelname.}
\vspace{2mm}
\label{tab:ablation-feature}
\begin{tabular}{lcc}
\toprule
Models & MRR & Delta \\
\midrule
\modelname$_\text{uniform}$ & $0.0376$ & \textcolor{gray}{-} \\
\quad remove Playlist Token & $0.0281$ & \textcolor{red}{$-0.0096$} \\
\quad remove Metadata Token & $0.0337$ & \textcolor{red}{$-0.0039$} \\
\quad remove Attribute Token & $0.0372$ & \textcolor{red}{$-0.0005$} \\
\quad remove Lyrics Token & $0.0374$ & \textcolor{red}{$-0.0003$} \\
\quad remove Audio Token & $0.0407$ & \textcolor{blue}{$+0.0036$} \\
\bottomrule
\end{tabular}
\end{wraptable}

The leave-one-out experiments presented in Table~\ref{tab:ablation-feature} provide further empirical support for this finding. The ablation of playlist tokens results in the most significant performance degradation ($-0.0096$ MRR), followed by metadata tokens ($-0.0039$ MRR). This hierarchical impact aligns with our theoretical understanding that playlist co-occurrence and metadata capture essential contextual signals about music similarity and user preferences. A particularly noteworthy observation is that removing audio tokens leads to a slight performance improvement ($+0.0036$ MRR). 
This counter-intuitive result suggests that the low-level audio features used in \modelname may actually introduce noise in conversational recommendation scenarios, where user preferences are predominantly articulated through natural language rather than acoustic similarity.

These empirical findings carry important implications for the design of future conversational music recommendation systems. First, they demonstrate that diverse multimodal representations can enhance recommendation performance. While different features may contribute differently depending on the conversation context, our results suggest that coarse-grained features such as playlist and metadata tokens play a crucial role in recommendation quality. As a future direction, we suggest analyzing the relative importance of different modalities based on specific conversation characteristics to further optimize recommendation systems.

\section{Related Works}
\vspace{-3mm}

\textbf{Conversational Recommendation}: Conversational recommendation systems provide recommendations through natural language dialog with users~\cite{goker2000adaptive, christakopoulou2016towards, zhang2018towards}. Traditional approaches use separate modules for dialogue and recommendation~\cite{zhang2018towards, chen2019towards, zhou2020improving}, resulting in a complex pipeline. Recent research explores using LLMs to unify these components. Early studies investigated zero-shot recommendation capabilities through prompt engineering~\cite{he2023large, liu2023chatgpt}, demonstrating LLMs could generate reasonable recommendations without specific training. However, a key challenge is the semantic gap between LLMs and recommender systems, as items are typically indexed by unique identifiers outside the LLM's vocabulary~\cite{hua2023tutorial}, making direct application difficult. More recent approaches address this by fine-tuning LLMs on recommendation data~\cite{zheng2024adapting, zhang2023recommendation, geng2023vip5}, enabling them to learn associations between natural language queries and item representations.

\textbf{Generative Retrieval}: Generative retrieval reformulates traditional retrieval as a generation problem, directly generating item identifiers instead of using similarity matching. This approach offers significant advantages over traditional approaches, including constant computational complexity regardless of database size and improved performance through structured item identifier generation. Differentiable Search Index (DSI)~\cite{tay2022transformer} pioneered generative retrieval, predicting document identifiers directly from queries using transformers without explicit similarity computations. Neural Corpus Indexer (NCI)~\cite{wang2022neural} built on this, generates document identifiers directly from queries using hierarchical semantic encoding and a prefix-aware weight-adaptive decoder. Recently, Transformer Index for Generative Recommenders (TIGER)~\cite{rajput2023recommender} developed a generative retrieval approach for recommendation similar to the proposed system, directly generating content-based item tokens using semantic ID generation through content feature encoding and Residual Vector Quantization (RVQ).

\textbf{Music Retrieval and Recommendation}: Music recommendation systems face unique challenges due to complex musical features and user behaviors~\cite{schedl2018current}. Most of the traditional approaches fall into three categories: collaborative filtering~\cite{hu2008collaborative}, content-based methods~\cite{van2013deep, knees2013survey, oramas2017deep}, and hybrid approaches combining both strategies. Natural language-based music retrieval has emerged to address limitations of traditional recommendation systems, enabling users to express preferences through text~\cite{manco2022contrastive, huang2022mulan, doh2023toward, doh2024enriching}. Conversational music recommendation systems extend this approach through multi-turn dialogues for more interactive discovery~\cite{chaganty2023beyond, doh2024music}. However, these systems typically separate dialogue management from recommendation components, using language models as an human interface. Developing a unified solution that learns the end-to-end connection of user queries, recommendations, system response in a multi-turn manner remains an open challenge in the field.

To address these challenges, \modelname employs an LLM as a conversational music recommendation engine for both item retrieval and the dialogue management. While we propose a generative retrieval approach similar to ~\cite{text2tracks}, our research differs significantly by being an LLM-based conversational system that incorporates a multimodal tokenizer combining not only metadata but also audio features, lyrics, playlist co-occurrence information, and semantic tokens.

\vspace{-3mm}
\section{Conclusion}
\vspace{-3mm}

We presented \modelname, demonstrating that music recommendation can be effectively reformulated as a token generation task within a large language model (LLM) framework. This novel perspective simplifies the architecture of recommendation systems by unifying dialogue understanding and recommendation generation into a single model. Our key contributions, the generation of synthetic conversations for training, a multimodal tokenization strategy, and a unified conversational recommendation model—collectively enable stable end-to-end training while delivering competitive performance. Notably, this approach reduces system complexity by eliminating the need for separate components such as dialogue managers or retrieval-based ranking modules. Empirical results indicate that \modelname maintains high recommendation quality across different modalities, including text and audio, highlighting its generalizability and robustness.

Looking ahead, there are several promising directions for future research. Scaling the model to larger and more diverse datasets could further enhance its recommendation accuracy and cultural relevance. Incorporating additional modalities—such as visual content or user behavior logs—may improve contextual understanding and personalization. Moreover, leveraging LLM-specific capabilities, such as chain-of-thought reasoning, zero-shot generalization, and in-context learning, offers exciting opportunities to push the boundaries of conversational recommendation systems. 

{\small
\bibliography{reference}
\bibliographystyle{abbrv}
}

\newpage
\appendix
\onecolumn

\section{Synthetic Data} \label{app:synth}
\subsection{Prompt to Google Gemini to synthesize conversation}\label{subapp:synth-prompt}

% after adding 5127 more,
% which is 2 bom, eom
% then 1025 x 5 tokens
% including -1 index 
% new token size: 133384

\begin{lstlisting}[style=docstring]
You are an AI assistant specializing in music. Generate an ANSWER to the QUESTION in valid LIST OF JSON:
- role: one of {"user", "assistant", "music"}
- content: 
  - for "user": string of conversation text for music request
  - for "assistant": string of conversation text for answer to user's request.
  - for "music": string of unique track_id (i.e., "3pzjHKrQSvXGHQ98dx18HI", "7vFv0yFGMJW3qVXbAd9BK9")

1. Conversation Start:
- The conversation begins with a user query.
  - This query can be a specific music item request (i.e., "superstition by steve wonder")
  - This query can be a natural language request (i.e., "party anthem with energetic rhythm", "play some 90s rock music")
- The assistant should respond with two consecutive objects:
  - A text response (role: "assistant") providing a brief reason for the recommendation
  - A music recommendation (role: "music") containing the `track_id`

2. Assistant Recommendations:
 - The assistant should recommend just one song per single turn, one that shares common musical traits (i.e, genre, artist, mood, theme, tempo) with the query song.
 - The goal is to syntehsize an engaging conversation and allow the user to explore music based on their request.
 - The assistant's music recommendations prioritize music tracks and tags, and ignore non-musical words. If there are captions and lyrics, use them as optional information as needed.

3. User Satisfaction:
 - In each turn, the user may or may not be satisfied with the recommendations and simply listen.
 - The user may express agreement with the recommendation (e.g., "This is great!").

4. User Direction Change:
 - In the real-world scenario, sometimes users are not happy with the recommendation. We need to include it in our synthetic conversation. 
 - In such a case, the user would express dissatisfaction with a recommendation.
 - The user may specify a new direction or trend of the recommended songs. (e.g., "Play something different", "Play something more upbeat").
 - Use the songs in the "NEGATIVE SET" to synthesize this recommendation that user does not like.
 - This helps evaluate how the assistant responds to negative feedback and adapts its recommendations.

5. assistant Adaptation:
The assistant should provide the next music track based on the following factors:
 - The previous conversation.
 - The current music being played.
 - The user's recent requests.
 - The assistant should recommend at least half of the songs in the provided playlist.

7. JSON Format:
 - The conversation should be formatted as a list of JSON objects.
 
8. Example:
QUESTION : 
Music Playlist:
{'5tkDCf3xKRZjT5RbCb3sZB': "Metadata:\n  Title: Big Amount\n  Artist: 2 Chainz, Drake\n  Album: Pretty Girls Like Trap Music, Industry Kings 2\nTag: drake, Hip-Hop/Rap, rap\n  Popularity: 55.0\nLyrics:  Pat downs, [REST OF THE LYRICS ARE OMITTED]\nPseudo Caption: This music features a hip hop song that consists of a flat male vocal rapping over shimmering hi hats, claps, punchy kick, groovy bass, synth lead melody and tinny hi hats. It sounds energetic, exciting and addictive - like something you would hear in clubs.\nTempo Estimate: 134.13\nKey Estimate: G# minor", 
[REST OF THE TRACKS ARE OMITTED]}

Negative Set:
{'7bJ4mu7MHa3rHiNyKjOoSl': 'Metadata:\n  Title: Ice Melts\n  Artist: Young Thug, Drake\n  Album: More Life\nTag: young thug, Hip-Hop/Rap\n  Popularity: 60.0\nLyrics:  You need to give it up to someone. [REST OF THE LYRICS ARE OMITTED]\nPseudo Caption: This music features a hip hop song that consists of a passionate male vocal, alongside harmonizing background vocals, singing over shimmering hi hats, punchy kick and claps. It sounds groovy and addictive - like something you would hear in clubs.\nTempo Estimate: 124.51\nKey Estimate: A major', 
[REST OF THE TRACKS ARE OMITTED]}

ANSWER:
[
  {"role": "user","content": "Play something like 'Big Amount' by 2 Chainz and Drake"},
  {"role": "assistant","content": "Here's a similar track with a similar vibe, featuring a catchy hook and hard-hitting flow."},
  {"role": "music","content": "4ckuS4Nj4FZ7i3Def3Br8W"},
  {"role": "user","content": "This is great! Can you play something more upbeat, but still in a hip hop style?"},
  {"role": "assistant","content": "Here's a track with a faster tempo and energetic rhythm, perfect for keeping the party going."},
  {"role": "music","content": "0B8QzDH7YWih85V5SEMnyJ"},
  {"role": "user","content": "I'm not feeling this one, play something different"},
  {"role": "assistant","content": "I understand. How about something with a bit more of a soulful feel, but still with a good groove?"},
  {"role": "music","content": "0PXp9QxBfSKGGpWUNaws2X"},
  {"role": "user","content": "Yeah, that's better, this is more my style"},
  {"role": "assistant","content": "Great choice! Let's keep it in this vein, with a focus on smooth vocals and melodic rhythms."},
  {"role": "music","content": "0H8XeaJunhvpBdBFIYi6Sh"},
  {"role": "user","content": "This is so good, this mood is perfect"},
  {"role": "assistant","content": "I'm glad you're enjoying it! Here's another track that leans into the same kind of soulful vibe."},
  {"role": "music","content": "3LwYGYpsg5SQioMX5BwFAz"},
  {"role": "user","content": "This is really good! One last request, something a little more intense with a little more aggression"},
  {"role": "assistant","content": "You got it. Here's a track with a harder edge and more energetic production. Get ready to bump!"},
  {"role": "music","content": "0utlOiJy2weVl9WTkcEWHy"}
]
\end{lstlisting}

\subsection{Synthetic Conversation Example}\label{subapp:synth-convo}

This is one of the synthetic conversation results. 
\begin{lstlisting}[style=json]
[{"role": "user", "content": "I'm looking for some mellow acoustic tracks to relax to. Can you suggest something?"},
 {"role": "music", "content": "3kZC0ZmFWrEHdUCmUqlvgZ"},
 {"role": "assistant", "content": "I think I Will Follow You into the Dark by Death Cab for Cutie would suit your mood. It's known for its emotional depth and gentle acoustic sound."},
 {"role": "user", "content": "That was beautiful, but a little too somber.  How about something with more energy, still in the alternative rock genre?"},
 {"role": "music", "content": "3mcG2NI5G5vhrQtRda1YnA"},
 {"role": "assistant", "content": "Okay, let's try something more upbeat. The Middle by Jimmy Eat World is a popular alternative rock anthem that offers a good balance of energy and emotion."}]
\end{lstlisting}

% \subsection{After Music Tokenization}\label{subapp:synth-final}

% This is after applying the music tokenization process in Section~\ref{subsec:system-music-token}. The music identifiers are converted to a sequence of multimodal music tokens. This is the text data fed to the LLM.

% \begin{lstlisting}[style=json]
% "<|begin_of_text|><|start_header_id|>system<|end_header_id|>\n\nCutting Knowledge Date: December 2023\nToday Date: 20 Jan 2025\n\nYou are a music expert. Your task is conversational retrieval: Predict the next music item based on the given chat history.<|eot_id|>
% <|start_header_id|>user<|end_header_id|>\n\nI'm looking for some mellow acoustic tracks to relax to.  Can you suggest something?<|eot_id|><|start_header_id|>assistant<|end_header_id|>\n\n<start_of_music><|playlist-53|><|semantic-559|><|metadata-895|><|lyrics-654|><|audio-268|><end_of_music>I think I Will Follow You into the Dark by Death Cab for Cutie would suit your mood. It's known for its emotional depth and gentle acoustic sound.<|eot_id|><|start_header_id|>user<|end_header_id|>\n\nThat was beautiful, but a little too somber.  How about something with more energy, still in the alternative rock genre?<|eot_id|><|start_header_id|>assistant<|end_header_id|>\n\n<start_of_music><|playlist-86|><|semantic-165|><|metadata-767|><|lyrics-158|><|audio-743|><end_of_music>Okay, let's try something more upbeat. The Middle by Jimmy Eat World is a popular alternative rock anthem that offers a good balance of energy and emotion.<|eot_id|>"
% \end{lstlisting}

\section{A-vs-B Human Evaluation Interface}~\label{appendix_a_3}

The Survey Interface is a web-based evaluation tool designed to allow participants to compare outputs from two models (A and B). For each evaluation item, users are presented with the same input prompt along with two model responses shown in random order. Participants are asked to choose the better response based on specific criteria—such as Recommendation Relevance and Response Naturalness—or select "Tie".

\begin{figure}[h]
\centering
\includegraphics[width= \linewidth]{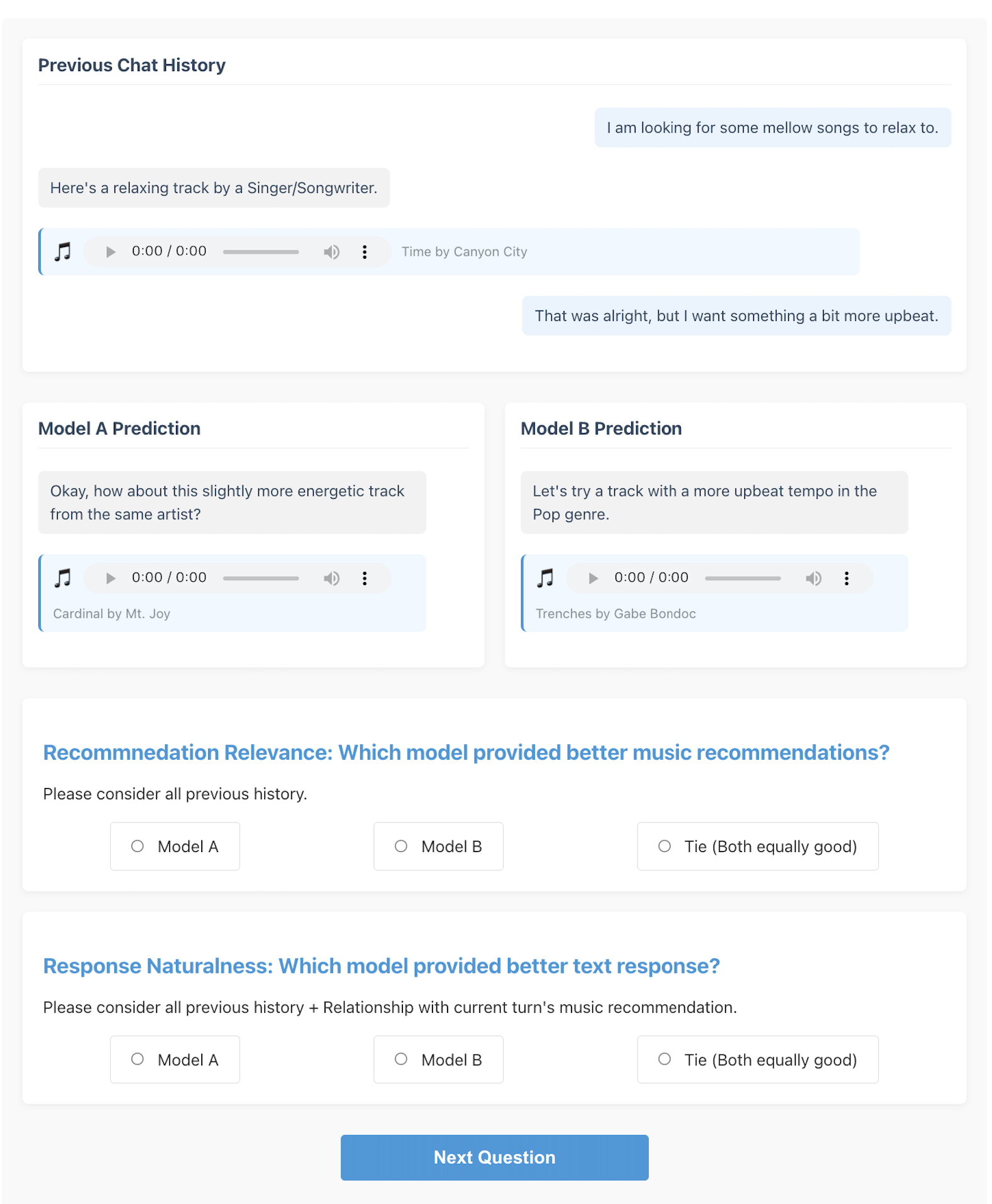}
\caption{Survey Interface of Subjective Evaluation.}
\label{fig:survey}
\end{figure}

\newpage

\section{Comparison Music Recommendation Models}~\label{appendix_a_4}

We compare our LLM-based conversational recommender system, \modelname, against five representative baselines that span different recommendation frameworks (sparse, similarity-based, generative retrieval) and item representation approaches (text, audio, listening history), each of which represents a distinct methodology for recommendation:

\textbf{BM25}~\cite{robertson1999okapi} represents a classical text-based sparse retrieval approach. It utilizes term frequency-inverse document frequency (TF-IDF) scoring to match user queries with text-based item representations. While computationally efficient and interpretable, this method relies solely on lexical matching and cannot capture semantic relationships or cross-modal patterns.

\textbf{NV-Embeds-V2}~\cite{lee2024nv} advances beyond sparse retrieval by implementing a text-based deep similarity retrieval framework. The model learns dense neural embeddings for both queries and music descriptions, enabling semantic matching that can capture latent relationships beyond exact word matches. This model represents a strong baseline as an LLM-based embedding extractor model that has demonstrated superior retrieval performance on the MTEB benchmark.

\textbf{CLAP}~\cite{wu2023large} extends the similarity-based paradigm to multiple modalities by jointly embedding audio and text. Through contrastive learning, it aligns the representations of music audio with corresponding textual descriptions, enabling cross-modal retrieval. This model represents the current state-of-the-art in audio-text music retrieval, providing a strong baseline for multimodal understanding.

\textbf{SASRec}~\cite{kang2018self} represents a sequential recommendation approach based on self-attention mechanisms. The model treats each music track as a unique token and learns to predict the next item in a sequence using transformer-based architecture. While effective for capturing sequential patterns in user listening history, it lacks the ability to process natural language queries or content information, making it primarily suitable for next-item prediction based on historical interactions. For the architecture in the default version of SASRec, we use two self-attention blocks with 50 latent dimensions. The model is trained using the Adam optimizer with a learning rate of 0.001, batch size of 128, dropout rate of 0.2, and maximum sequence length of 50.

\textbf{TIGER}~\cite{rajput2023recommender} implements a generative retrieval framework that combines semantic understanding with residual vector quantization. Instead of using unique item IDs, it represents each music track using Semantic IDs composed of SentenceT5 embeddings and residual vector quantization (RVQ). The model is implemented using a transformer-based encoder-decoder architecture containing 4 layers each with 6 self-attention heads, using a vocabulary of 1024 semantic tokens (256 x 4). The model is trained with a batch size of 256, using a learning rate of 0.01.

For text-based content retrieval models (BM25, NV-Embeds-V2, TIGER), item representations include metadata and semantic tags,~\footnote{\{Title\} by \{Artist\} from \{Album\} \{Semantic tags\} \{Year\}} excluding lyrics, playlist context, and audio modalities. We initially included lyrics in our item representations but found that this led to significant performance degradation. Therefore, we excluded lyrics from the final item representations. CLAP, being an audio-text joint embedding model, restricts item representations to only the audio modality, while SASRec, as a sequential recommendation model, represents items using only random IDs without any content information.

\end{document}